\documentclass[%
 reprint,
  superscriptaddress,
 amsmath,amssymb,amsfonts,
 aps,
pra,
 floatfix,
 longbibliography
]{revtex4-2}

\usepackage[utf8]{inputenc}
\usepackage[pdftex]{graphicx} \graphicspath{{}}
\usepackage{float} 
\usepackage{ulem}
\usepackage{color}
\usepackage[dvipsnames]{xcolor}
\usepackage[pdftex,colorlinks=true]{hyperref}
\hypersetup{
  colorlinks=true,
  linkcolor=blue,
  citecolor = blue,
  urlcolor=blue,
}

\usepackage{tikz}
\newcommand*\emptycirc[1][0.4ex]{\tikz\draw (0,0) circle (#1);} \newcommand*\fullcirc[1][0.4ex]{\tikz\fill (0,0) circle (#1);} 

\usepackage{mathtools}

\DeclarePairedDelimiterX\braket[2]{\langle}{\rangle}{#1 \delimsize\vert #2}
\DeclarePairedDelimiterX\expval[3]{\langle}{\rangle}{#1 \delimsize\vert #2  \delimsize\vert #3}
\DeclarePairedDelimiterX\proj[2]{\delimsize\vert#1\rangle}{\langle#2\delimsize\vert}{ }
\usepackage{siunitx}
\begin{document}

\title{Polarons and bipolarons in Rydberg-dressed extended Bose-Hubbard model}
\author{G. A. Dom\'inguez-Castro}
\affiliation{Centro de Nanociencias y Nanotecnolog\'ia, Universidad Nacional Aut\'onoma de M\'exico, Apartado Postal 14, 22800 Ensenada, Baja California, M\'exico}
\author{L. Santos}
\affiliation{Institut f\"ur Theoretische Physik, Leibniz Universit\"at Hannover, Appelstrasse 2, D-30167 Hannover, Germany}
%\email[]{Your e-mail address}
%\homepage[]{Your web page}
%\thanks{}
%\altaffiliation{}
\author{L. A. Pe\~na Ardila}
\affiliation{School of Science and Technology, Physics Divisio, University of Camerino, Via Madonna delle Carceri, 9B - 62032 (MC), Italy}
\affiliation{Dipartimento di Fisica,Università di Trieste, Strada Costiera 11, I-34151 Trieste, Italy}
\email{luisaldemar.penaardila@units.it}
\date{\today}

\begin{abstract}
Impurities immersed in hard-core Bose gases offer exciting opportunities to explore polaron and bipolaron physics. We investigate the ground state properties of a single and a pair of impurities throughout the superfluid and insulating (charge density wave) phases of the bosonic environment. In the superfluid phase, the impurity exhibits polaron-like behavior, forming a dressed quasiparticle. In contrast, in the insulating phase, the impurity regains its particle-like character, moving through a potential landscape shaped by the charge density wave order. Moreover, we show that two impurities can form a bound state even in the absence of an explicit impurity-impurity coupling. We establish the stability of this bound state within both the superfluid and insulating phases. Our results offer valuable insights for ongoing lattice polaron experiments with ultracold gases.
\end{abstract}

\maketitle

\section{Introduction}

Polarons arising from the interaction between mobile impurities and their host environment have provided valuable insights into quantum many-body systems~\cite{landau1948effective,pekar1958theory}, offering simple yet accurate descriptions of a wide range of physical systems, including liquid helium mixtures~\cite{baym2008landau,PhysRev.156.207}, ultracold atomic gases~\cite{scazza2022repulsive,massignan2014polarons,chevy2010ultra,Schmidt_2018_REV,Ardila_Review}, doped antiferromagnets~\cite{PhysRevB.40.6721,PhysRevB.39.6880}, and hybrid light-matter materials~\cite{sidler2017fermi,takemura2014polaritonic}, among others. On the other hand, ultracold atoms in optical lattices provide a powerful quantum simulation platform for strongly correlated phenomena~\cite{Bloch2012,Georgescu14}. Interestingly due to the translational breaking of symmetry, new phases arise when optical lattices are doped as the size of the polaron can be comparable with the lattice size
~\cite{leskinen2010fermi,PhysRevA.93.053601,PhysRevA.76.011605,keiler2020doping,dominguez2023bose}. This includes the behavior of strongly interacting polarons in environments undergoing quantum phase transitions \cite{PhysRevA.107.063309,PhysRevLett.130.173002,PhysRevA.105.063303}, impurity dynamics \cite{fukuhara2013quantum,massel2013dynamics,PhysRevA.82.063614,bruderer2008transport,PhysRevB.93.125110,PhysRevA.95.063605,PhysRevA.110.063306}, the formation of bipolarons \cite{PhysRevA.105.L021303,PhysRevB.100.245419,PhysRevA.88.053601,ding2023polarons,yordanov2023mobile,isaule2024bound,santiago2023collective,PhysRevA.110.030101,ArdilaMultiPol,Astrakharchik2023,gómezlozada2024}, polarons in topological media \cite{PhysRevB.99.081105,PhysRevB.103.245106,PhysRevB.100.075126,vashisht2024chiralpolaronformationedge,PhysRevB.109.195153}, and the intricate interplay between kinetic energy and spin interactions in doped Mott insulators \cite{koepsell2021microscopic,koepsell2019imaging,lebrat2024observation}. 
\begin{figure}[t!]
\centering
\includegraphics[width=0.99\columnwidth]{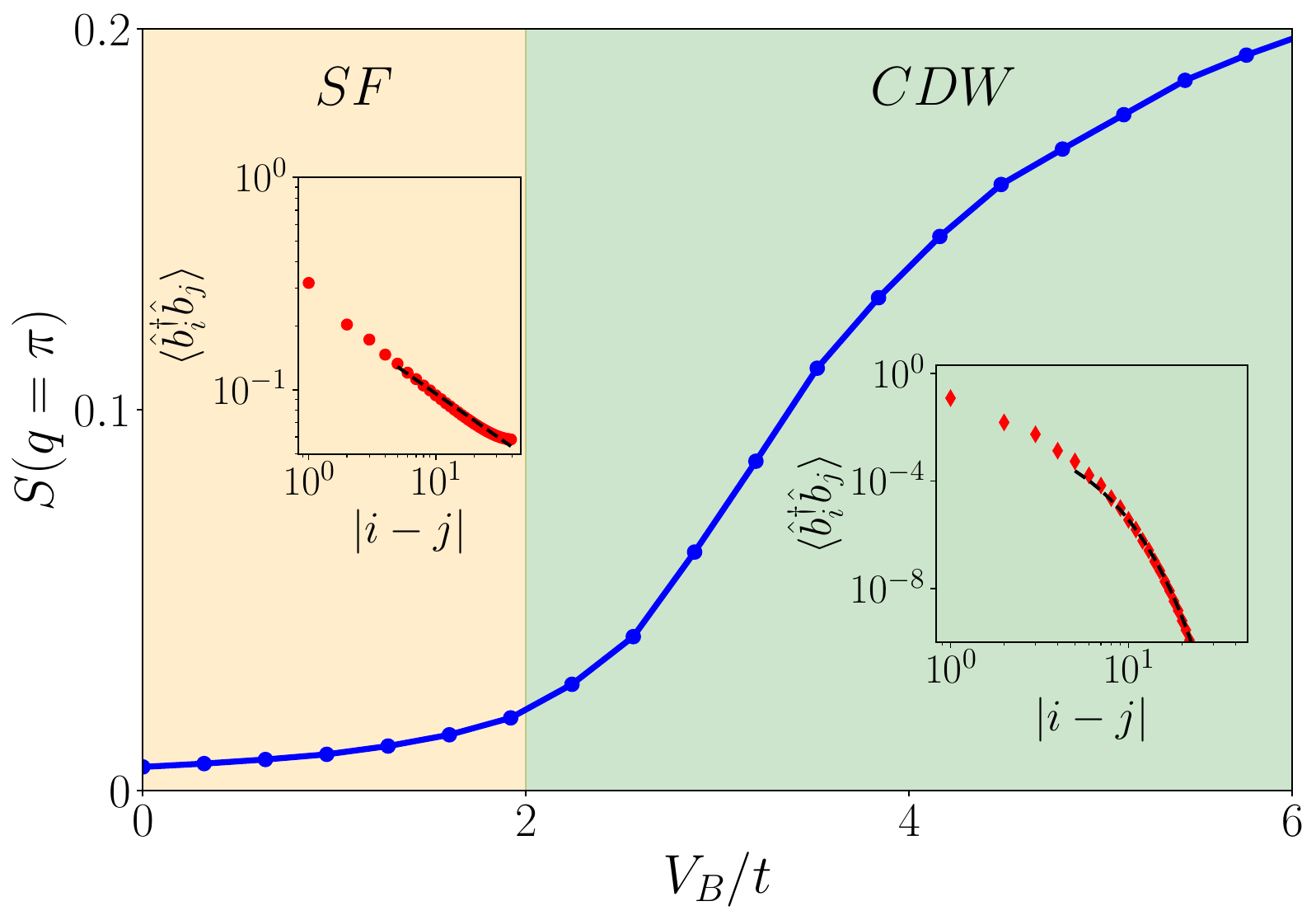}
\caption{Structure factor $S(q=\pi)$ of the bosonic medium as a function of the boson-boson interaction $V_{B}/t$. The insets display the characteristic one-body correlation functions $\langle\hat{b}^{\dagger}_{i}\hat{b}_{j}\rangle$ as a function of the distance $|i-j|$. Markers are associated with DMRG calculations, while dashed lines correspond to algebraic and exponential fits in the SF ($V_{B}/t=0$) and CDW ($V_{B}/t=4$) phases, respectively. We consider $L=80$ and $n_{B}=1/2$. Away from half-filling the bosonic bath remains in the superfluid phase.\label{fig:1}}
\end{figure}

Recently, the one-dimensional extended Bose-Hubbard model (eBHM) with nearest-neighbor interactions was successfully realized using Rydberg-dressed atoms \cite{weckesser2024}. Furthermore, the eBHM with longer-range interactions was implemented using magnetic atoms \cite{baier2016extended} and could potentially be realized with polar molecules in the upcoming generation experiments \cite{yan2013observation,li2023tunable}. These breakthrough experiments open up exciting new avenues for exploring impurity physics within an itinerant, long-range interacting quantum gas. Moreover, the eBHM exhibits a superfluid-to-insulator phase transition \cite{PhysRevB.80.174519,PhysRevA.107.063307}, making it an appealing environment for studying the behavior of impurities across this transition. Understanding the underlying physics of impurities in insulating media has gained increasing relevance \cite{10.21468/SciPostPhys.16.2.056,amelio2024polaron}, particularly in light of recent semiconductor experiments \cite{smolenski2021signatures,PhysRevX.11.021027,PhysRevB.106.L241104}.

In this work, we study the quasiparticle properties of a single and a pair of impurities immersed in an interacting hard-core Bose gas on a one-dimensional lattice. Such a system can be realized by doping with impurities a lattice Rydberg-dressed gas \cite{weckesser2024}. In contrast to Ref. \cite{leskinen2010fermi}, which focused on the case of a single impurity in a noninteracting polarized Fermi gas, we describe the ground-state properties of an impurity across the superfluid (SF) and charge density wave (CDW) phases of the medium. In the superfluid phase, the impurity is dressed by the bosonic particles, showing polaron-like behavior. However, in the CDW phase, the impurity retains its single-particle character but moves in a potential landscape shaped by the CDW order, which can be effectively described by an ionic Hubbard model \cite{PhysRevB.84.195102,PhysRevResearch.4.033119,PhysRevB.96.035116,PhysRevLett.92.246405,PhysRevB.63.235108} (IHM). 
For the case of two impurities, we numerically demonstrate that the impurities can form a bound state even in the absence of an explicit impurity-impurity coupling within both the SF and CDW phases of the medium. \\

This manuscript is organized as follows. We introduce the model under study in Sec. \ref{model} and briefly review the properties of the hard-core Bose gas in Sec. \ref{BM}. Afterward, in Sec. \ref{SI}, we examine the binding energy, quasiparticle residue, and resulting spatial bath-impurity correlations for a single impurity. Section \ref{TI} addresses the two-impurity problem, with a particular focus on bipolaron formation. Finally, in Sec. \ref{CO}, we summarize our findings.

\section{Model}
\label{model}
We consider mobile impurities doping interacting hard-core bosons in a one-dimensional lattice. The Hamiltonian of the system can be written as $\hat{H} = \hat{H}_{B}+\hat{H}_{I}+\hat{H}_{BI}$, where $\hat{H}_{B}$ describes the bosonic bath, $\hat{H}_{I}$ the impurities, and $\hat{H}_{BI}$ denotes the coupling between the bath and the impurities. Explicitly, each component is given as follows:
\begin{equation}
\begin{split}
\hat{H}_{B} &= -t_{B}\sum_{\langle i,j \rangle} \hat{b}_{i}^{\dagger}\hat{b}_{j} + V_{B}\sum_{i}\hat{n}_{B,i}\hat{n}_{B,i+1}\\
\hat{H}_{I} &=  -t_{I}\sum_{\langle i,j \rangle} \hat{a}_{i}^{\dagger}\hat{a}_{j} \ \ \
\hat{H}_{BI} = U_{BI}\sum_{i}\hat{n}_{I,i}\hat{n}_{B,i},
\end{split}
\label{Eq1}
\end{equation}
where operators $\hat{b}_{i}^{\dagger}$ ($\hat{b}_{i}$) and $\hat{a}_{i}^{\dagger}$ ($\hat{a}_{i}$) create (annihilate) a
boson and an impurity, respectively, at lattice site $i$. The particle number operator for the bosons is $\hat{n}_{B,i} = \hat{b}_{i}^{\dagger}\hat{b}_{i}$, and for the impurities  $\hat{n}_{I,i} = \hat{a}_{i}^{\dagger}\hat{a}_{i}$. The nearest neighbor hopping amplitude is $t_{B}$ and $t_{I}$ for the bosons and impurities, respectively; $V_{B}$ is the nearest-neighbor repulsive interaction among the particles in the bath, and $U_{BI}$ is the on-site interaction between the bosons and the impurities. 
The hard-core constraint of the bath atoms forbids double occupancy, i.e. $(b_{i}^{\dagger})^{2}=0$. This restriction can be experimentally achieved using sufficiently strong on-site interactions \cite{weckesser2024}. We do not impose the hard-core constraint on the impurities. Notice that $\hat{H}_{I}$ does not explicitly include an impurity-impurity interaction term.

To investigate the ground state properties of a single and a pair of impurities, we employ density matrix renormalization group (DMRG) techniques using the TeNPy library \cite{TeNPy2018}. All calculations assume periodic boundary conditions on a lattice with $L$ = 80 sites and a maximum matrix product state (MPS) bond dimension of $\chi=800$. For concreteness, we consider the case where $t_{B}=t_{I}=t$. 
\begin{figure}[t!]
\centering
\includegraphics[width=0.99\columnwidth]{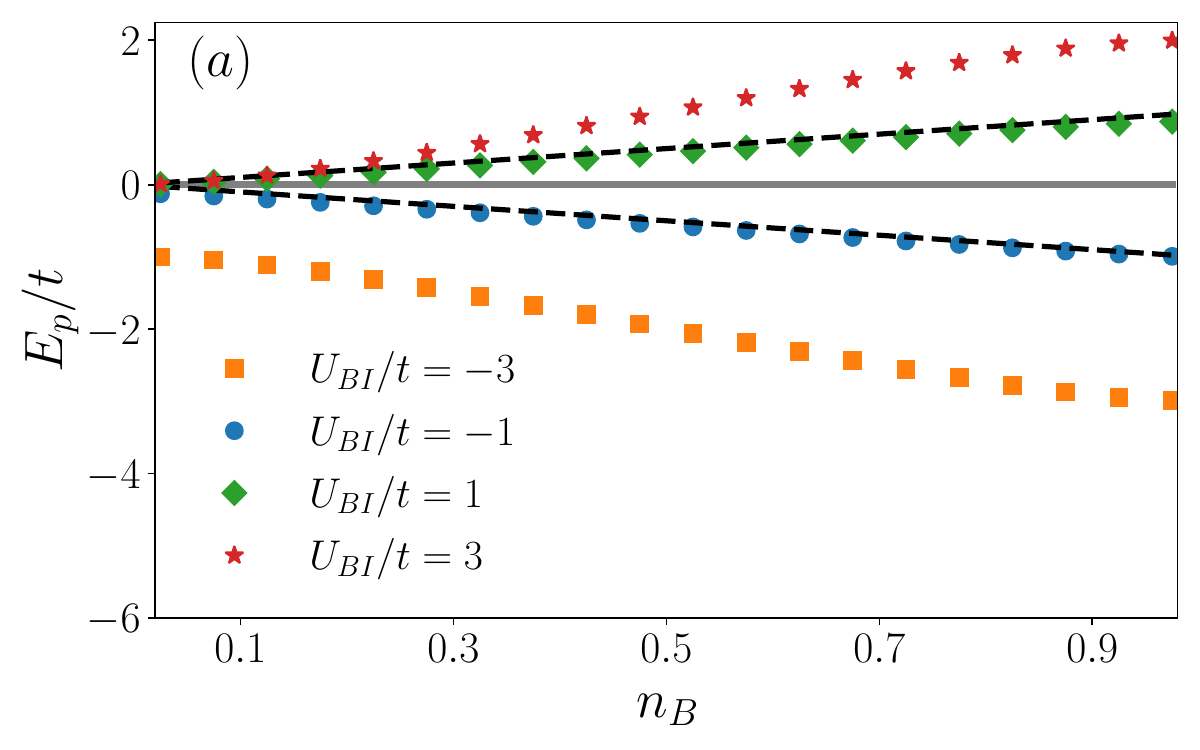}
\includegraphics[width=0.99\columnwidth]{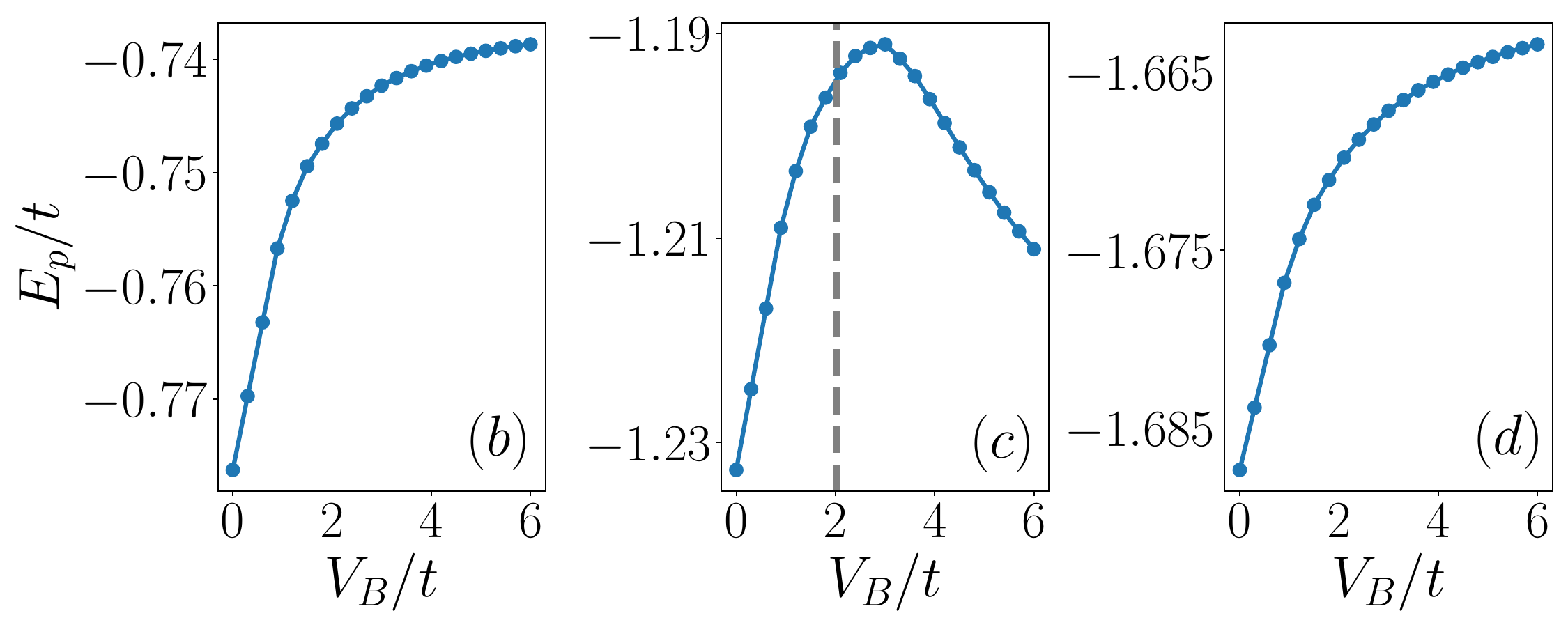}
\caption{(a) Binding energy $E_{p}$ of a single impurity as a function of the density of the bath $n_{B}$ for $V_{B}/t = 0$. Markers are associated with DMRG calculations, whereas the dashed lines correspond to the mean-field prediction $E_{p}\approx U_{BI}n_{B}$. The gray line represents a guide for the eye for zero energy. (b)-(d) Binding energy $E_{p}$ of a single impurity as a function of the boson-boson interaction $V_{B}/t$ for $n_{B} = 0.25, 0.5$ and $0.75$, respectively. The grey dashed line indicates the value of $V_{B}/t$ at which the SF-CDW transition takes place. We consider $U_{BI}/t=-2$.
\label{fig:2}}
\end{figure}

\section{Bosonic medium}
\label{BM}
Before delving into the study of impurities, we briefly review the ground-state properties of the bosonic medium in the absence of impurities. For repulsive inter-site interactions $V_{B}>0$, hard-core bosons exhibit either a superfluid or an insulating charge density wave phase. The latter occurs exclusively at half filling $n_{B}=1/2$ and for $V_{B}/t>2$ \cite{giamarchi2003quantum}. As shown in Fig. \ref{fig:1}, the emergence of the CDW can be tracked by the structure factor $S(q)=\frac{1}{L^{2}}\sum_{j,\ell}e^{-iq(j-\ell)}\langle\hat{n}_{B,j}\hat{n}_{B,\ell}\rangle$, where $q\in [-\pi,\pi]$ is the quasi-momentum. The non-zero value of $S(q=\pi)$ around $V_{B}/t\approx 2 $ signals the formation of a density modulation,  which for $t = 0$ acquires the simple form $|\cdots\,\fullcirc\,\emptycirc\,\fullcirc\,\emptycirc\,\fullcirc\,\emptycirc\,\cdots\rangle$ ($\fullcirc$ and $\emptycirc$ denoting occupied and empty sites).

The insets of Fig. \ref{fig:1} illustrate the one-body correlation functions $\langle\hat{b}^{\dagger}_{i}\hat{b}_{j}\rangle$ within the SF and CDW phases. In the gapless SF phase, off-diagonal quasi-long-range order yields a power-law decay $\langle\hat{b}^{\dagger}_{i}\hat{b}_{j}\rangle \propto 1/|i-j|^{\alpha}$, whereas in the gapped CDW
it decays exponentially $\langle\hat{b}^{\dagger}_{i}\hat{b}_{j}\rangle \propto e^{-|i-j|/\xi}$. The SF supports phononic excitations, while in the strong-coupling CDW ($V_{B}/t \gg 1$) the lowest excitations are defects breaking the density-wave order.

\section{Single Impurity}
\label{SI}

We begin the single impurity analysis by computing the binding energy $E_{p}$, which corresponds to the energy required to add a single impurity into the medium. The binding energy can be computed as follows
\begin{equation}
E_{p} = E_{1}-(E_{0}+E_{I}),
\label{Eq2}
\end{equation}
where $E_{0}$ and $E_{1}$ denote the ground-state energies with zero and one impurity, respectively, and $E_{I} = -2t$ is the lattice dispersion energy of a single free impurity.

Figure  \ref{fig:2}(a) shows $E_{p}$ as a function of the medium density for several bath-impurity interactions at fixed $V_{B}/t = 0$. For weak couplings $U_{BI}\lesssim t$, $E_{p}$ follows the mean-field predictions, $E_{p}\approx U_{BI}n_{B}$ (black dashed lines). Stronger boson-impurity interactions reveal nonlinear density behavior, highlighting the relevance of beyond mean-field terms \cite{santiago2024lattice}. In the dilute limit of the bath and for attractive interactions $U_{BI}<0$, $E_{p}$ approaches the boson-impurity bound state energy $E_{p}=4t-\sqrt{|U_{BI}|^{2}+16t^{2}}$ \cite{KORNILOVITCH2024169574}. This is not the case for $U_{BI}>0$, as the repulsively bound pair is not the two-body ground state. Figures \ref{fig:2}(b)-\ref{fig:2}(d) display $E_{p}$ as a function of $V_{B}$ for three bosonic fillings and fixed $U_{BI}/t=-2$. For $n_{B}=0.25$ and $n_{B}=0.75$, $E_{p}$ increases monotonically with the boson-boson interaction. At half-filling, the binding energy bends near the superfluid-insulator transition, signaling a qualitative change in impurity behavior across the second-order phase transition of the medium.

To delve into the impurity behavior across the phase transition of the medium, we compute the quasiparticle residue $Z$ of the impurity. This quantity measures the spectral weight of the quasiparticle, quantifying how particle-like the impurity remains in an interacting environment. It is defined as the square of the overlap between the interacting ground state and the non-interacting boson-impurity state
\begin{equation}
Z = |\langle\Psi_{1}(n_{B},V_{B},U_{BI})|\Psi_{1}(n_{B},V_{B},0)\rangle|^{2},
\label{Eq3}
\end{equation}
where $\Psi_{1}(n_{B},V_{B},U_{BI})$ denotes the ground state of the environment with one impurity for the system parameters $(n_{B},V_{B},U_{BI})$.

\begin{figure}[t!]
\centering
\includegraphics[width=1.0\columnwidth,height=0.78\columnwidth]{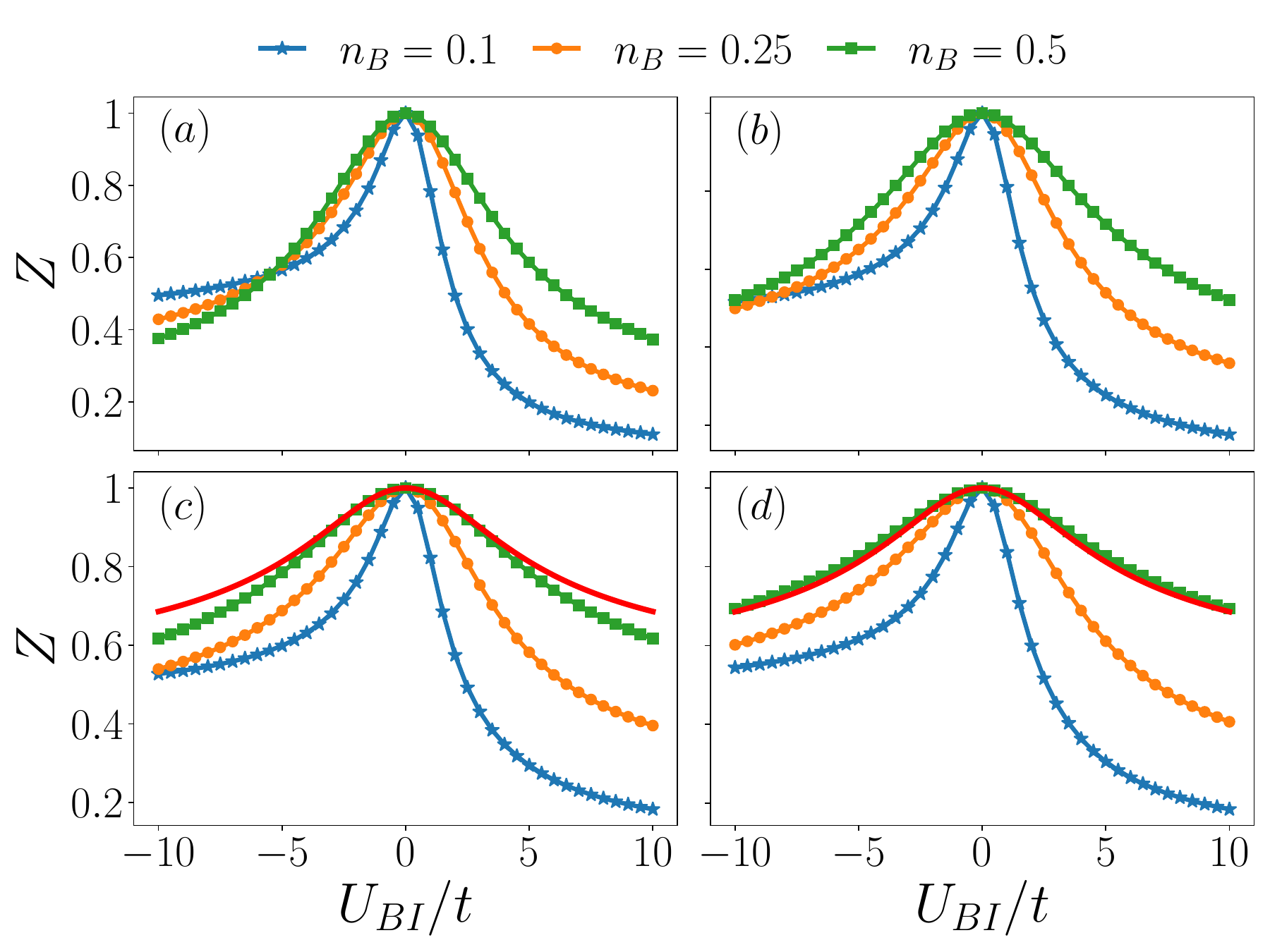}
\caption{Quasiparticle residue $Z$ as a function of the boson-impurity interaction $U_{BI}/t$. Panels (a), (b), (c), and (d) are associated with $V_{B}/t = 0,1,2$, and $6$, respectively. The red line in panels (c) and (d) corresponds to the overlap between the unmodulated and modulated single-particle ground states of the ionic Hubbard model. 
\label{fig:3}}
\end{figure}

Figures. \ref{fig:3}(a)-\ref{fig:3}(d) show $Z$ as a function of $U_{BI}/t$ for various values of $n_{B}$ and $V_{B}/t$. Notice that for $n_{B}=0.1$ and $n_{B}=0.25$, the residue decreases faster for repulsive than for attractive couplings. This is because the impurity is dressed by holes in the repulsive branch and since the hole density is higher for $0<n_{B}<1/2$, the impurity becomes more strongly dressed by the medium, quickly losing its individual particle-like character. By particle-hole symmetry, the opposite occurs for $1/2<n_{B}<1$ (see Appendix \ref{Ap1}). The half filling case is symmetric since particle and hole densities are equal.

In the dilute limit of the bath $n_{B}=0.1$, the residue barely changes with $V_{B}$ on the attractive side. This is consistent with the picture that the impurity binds with a boson and the system behaves as a bound state moving in a sparse bosonic media. On the repulsive side, $Z$ decreases faster with increasing $V_{B}$ due to enhanced hole-hole correlations.  For $n_{B}=0.25$ and $n_{B}=0.5$ (away from the CDW transition), $Z\simeq 1$ for moderate boson-impurity couplings, consistent with a polaron-like regime~\cite{leskinen2010fermi} where the impurity behaves much like as a renormalized free particle. Stronger $U_{BI}/t$ couplings reduce $Z$, signaling the breakdown of the polaron picture as the spectral weight of the impurity spreads over many collective excitations.
\begin{figure*}[t!]
\centering
\includegraphics[width=2.0\columnwidth]{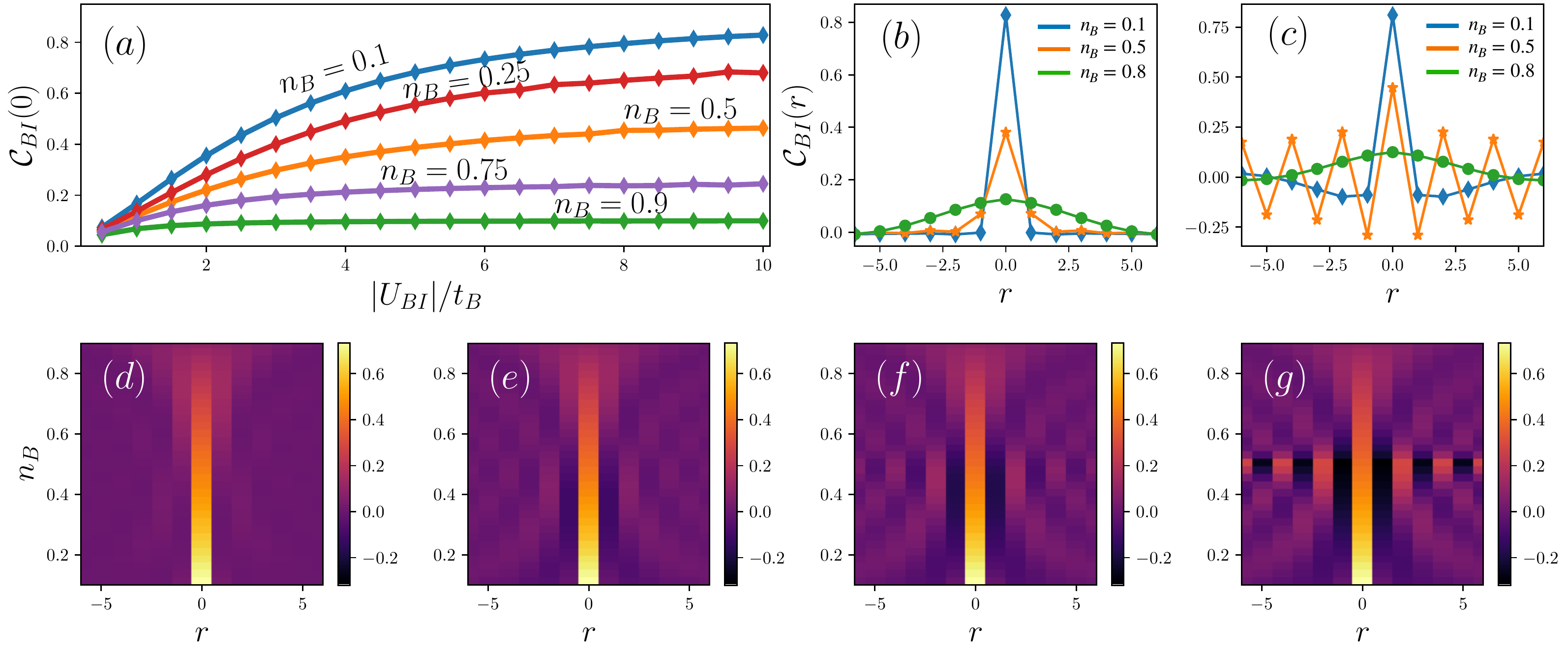}
\caption{(a) Density-density boson-impurity correlation $\mathcal{C}_{BI}(r)$ at the same site $(r=0)$ as a function of the interaction $U_{BI}/t$ for several bosonic filling factors $n_{B}$ and fixed $V_{B}=0$. (b) $\mathcal{C}_{BI}(r)$ as a function of the relative distance $r$ for three different values of $n_{B}$ and $(V_{B}/t,U_{BI}/t)=(0,-10)$. (c) Sames as panel (b) but for $(V_{B}/t, U_{BI}/t)=(4,-10)$. (d)-(g) Development of the density-density boson-impurity correlations as the medium transitions from the superfluid to the charge density wave for $U_{BI}/t =-6$. Panels (d), (e), (f) and (g) are associated with $V_{B}/t = 0,1,2$, and $4$ respectively.\label{fig:4}}
\end{figure*}

Surprisingly, as the bath enters the CDW phase, the residue increases, indicating that the impurity regains part of its free particle character. This occurs because the CDW phase features a gap of order $\propto V_{B}$, thus weak enough boson-impurity couplings do not significantly alter the CDW background. Consequently, the impurity moves on top of this static background, which behaves as an external potential. To illustrate the latter, let us consider the deep insulating regime of the bath, where density fluctuations can be neglected and the bosonic operators can be approximated by their expectation values $\langle\hat{b}_{i}\rangle\approx 0$ and $\langle\hat{n}_{B,i}\rangle\simeq 1/2[1+(-1)^i]$. This leads to the following single-particle Hamiltonian
\begin{equation}
\hat{H}_{I}^{\text{eff}} = -t\sum_{\langle i,j \rangle} \hat{a}_{i}^{\dagger}\hat{a}_{j}+\frac{U_{BI}}{2}\sum_{i}(-1)^{i}\hat{n}_{I,i},
\label{Eq4}
\end{equation}
where the boson-impurity coupling becomes an alternating local potential. The above approach is equivalent to considering the zero particle-hole excitations subspace of the CDW  \cite{amelio2024polaron}. The Hamiltonian in Eq.~\eqref{Eq4} takes the form of the celebrated ionic Hubbard model \cite{PhysRevB.84.195102,PhysRevResearch.4.033119,PhysRevB.96.035116,PhysRevLett.92.246405,PhysRevB.63.235108}. This model has been employed to describe the neutral-ionic transition in quasi-one-dimensional charge-transfer organic materials \cite{PhysRevLett.47.1750,strebel1970theory, nagaosa1986theory1, nagaosa1986theory2}, and the ferroelectric transitions in perovskites \cite{egami1993lattice, PhysRevLett.83.2014}. In our case, the potential modulation naturally arises from the interplay between the CDW order and the boson–impurity coupling. As depicted in Fig. \ref{fig:3}(c) and Fig. \ref{fig:3}(d), the square of the overlap between the unmodulated ($U_{BI}/t=0$) and modulated ($U_{BI}/t\neq0$) single-particle ground states of Eq.~\eqref{Eq4} matches with the DMRG calculations for $n_{B}=1/2$, proving the accuracy of the effective Hamiltonian.

To conclude the  single impurity section, we analyze the density-density boson-impurity correlation function
\begin{equation}
\mathcal{C}_{BI}(i-j) = L[\langle\hat{n}_{B,i}\hat{n}_{I,j} \rangle -\langle\hat{n}_{B,i}\rangle\langle\hat{n}_{I,j}\rangle],
\label{}
\end{equation}
which measures the deviation from the background value of the number of bosons at site $i$ given that the impurity is at site $j$. This correlator can be experimentally accessed via quantum gas microscopy or quench protocols~\cite{ArdilaPRL2018}. For simplicity, we focus on the attractive case $U_{BI}/t<0$, the results associated with $U_{BI}/t>0$ can be found in the Appendix \ref{Ap1}.

Figure \ref{fig:4}(a) shows $\mathcal{C}_{BI}(0)$ as a function of $U_{BI}/t$ for several values of $n_{B}$, we consider a noninteracting medium $V_{B}=0$. The local dressing initially grows linearly with $U_{BI}/t$, consistent with perturbative predictions~\cite{ding2023polarons}, and saturates at strong boson-impurity interactions to $\mathcal{C}_{BI}(0) \approx 1-n_{B}$, corresponding to the maximum density fluctuation of the medium. Qualitatively similar results are observed for different values of $V_{B}$. The nature of the environment becomes evident in the spatial behavior of $\mathcal{C}_{BI}(r=i- j)$. As shown in Fig. \ref{fig:4}(b), in the SF phase, the impurity is surrounded by an extended bosonic cloud on top and beyond $(r>0)$, in agreement with the polaron-like picture~\cite{leskinen2010fermi, ding2023polarons}. As the bath becomes strongly interacting, Fig. \ref{fig:4}(c), the cloud of bosons at $n_{B}=0.1$ and $n_{B}=0.5$ sharply contrast with the non-interacting case since bosons are pushed away from nearest neighboring sites of the impurity. Moreover, in the half-filled scenario, the impurity inherits the CDW correlations for large $U_{BI}$ since it binds with a boson. For $n_{B}=0.8$, the behavior is qualitatively similar for $V_{B}=0$ and $V_{B}=4t$ as the hard-core nature of the bath dominates the physics at high-densities.

Figures \ref{fig:4}(d)-\ref{fig:4}(g) display in a color scheme the development  of the boson-impurity correlations $\mathcal{C}_{BI}$ as the medium transitions from the superfluid to the CDW, we consider $U_{BI}/t=-6$. In the noninteracting case $V_{B}=0$, the bosons dress the impurity in the same fashion $\mathcal{C}_{BI}>0$ for all fillings. As $V_{B}$ increases the bosons repeal each other at neighboring sites leading to a proliferation of negative nearest-neighbor correlations. Interestingly, for $V_{B}/t=4$, CDW-like correlations are shown in a narrow band around half-filling. This is because ground states with fillings close to $n_{B}=0.5$ display weakly density arrangement as they can be viewed as particle or hole excitations of the half-filled ground state. However these CDW-like correlations vanish for small $r$ and truly inherited order happens for $n_{B}=0.5$ only.

\section{Two Impurities}
\label{TI}

We consider two soft-core bosonic impurities immersed in an interacting hard-core Bose gas.
The soft-core nature allows double occupancy of the impurities. To avoid the boson-impurity bound state, we focus on the repulsive branch $U_{BI}>0$ only. Although there is no explicit impurity–impurity interaction, density fluctuations in the surrounding bosonic bath can mediate an effective attraction between nearby impurities~\cite{ding2023polarons,PhysRevA.110.030101,PhysRevLett.121.013401,PhysRevLett.127.103401}. This interaction can be strong enough to bind two impurities, forming a bipolaron. To monitor bipolaron formation, 
we compute the binding energy of two impurities $E_{Bip}$:
\begin{equation}
E_{Bip} = E_{2}-2E_{1}+E_{0},
\label{}
\end{equation}
 where $E_{2}$, $E_{1}$, and $E_{0}$ denote the ground-state energies with two, one, and zero impurities, respectively. A bipolaron forms when $E_{Bip}<0$ \cite{note1}.

\begin{figure}[t!]
\centering
\includegraphics[width=1.0\columnwidth]{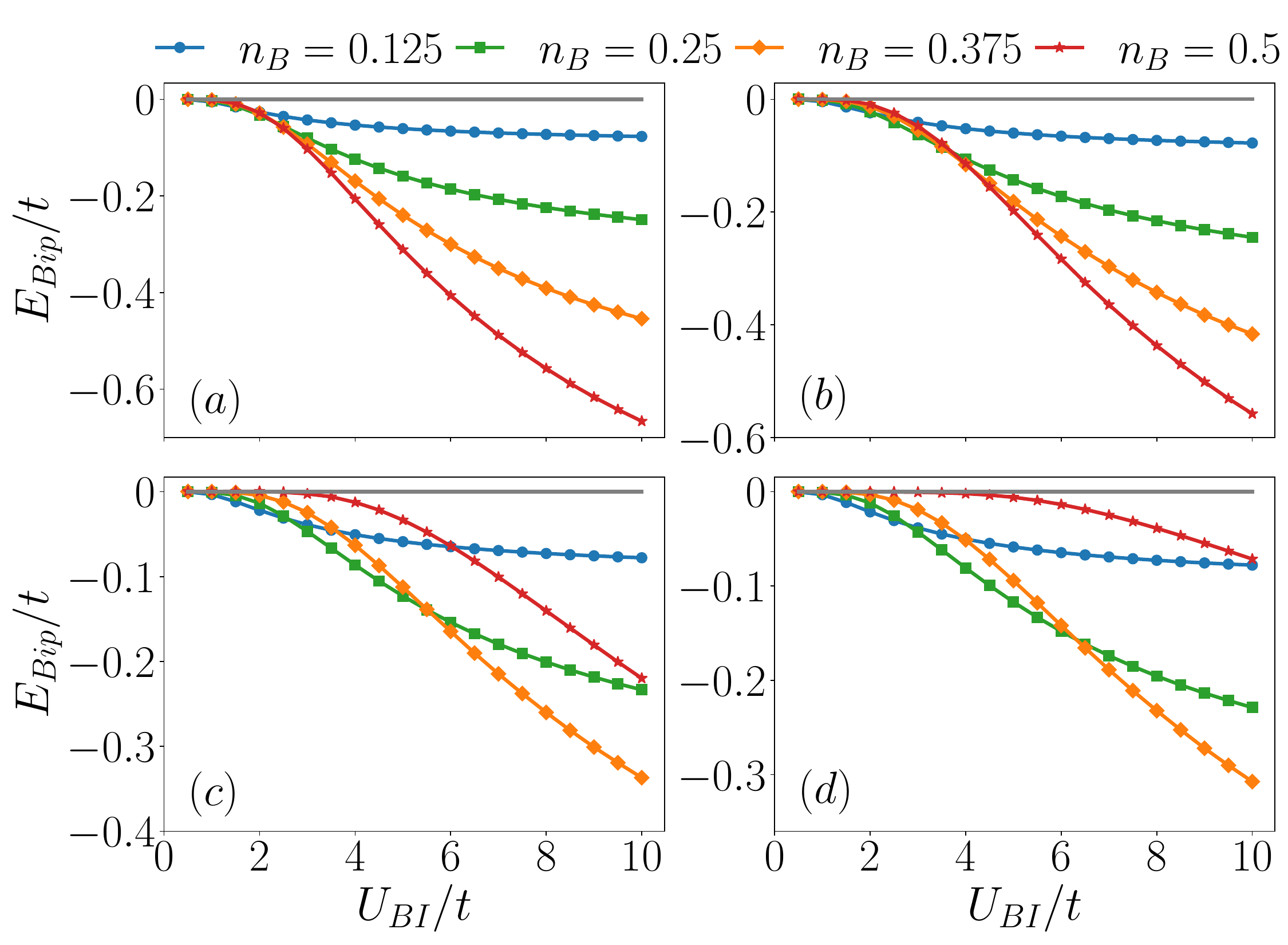}
\caption{Binding energy $E_{\mathrm{Bip}}$ of the two-impurity state as a function of the boson-impurity interaction $U_{BI}/t$ for several values of the bosonic filling $n_{B}$. Panels (a), (b), (c) and (d) consider $V_{B}/t= 0, 1, 3$ and $4$, respectively. The gray line represents a guide for the eye for zero energy.\label{fig:5}}
\end{figure}

Figures \ref{fig:5}(a)-\ref{fig:5}(d) illustrate $E_{Bip}$ as a function of the boson-impurity coupling for various bosonic fillings and boson-boson interaction strengths. In the noninteracting case $V_{B}=0$, the binding energy is nearly density independent at weak boson-impurity couplings, while stronger interactions lead to tighter binding in more compressible media, in agreement with bipolarons in homogeneous media \cite{PhysRevLett.121.013401}. As the bath becomes strongly interacting, density fluctuations decrease, hindering the binding of the impurities. This is clearly illustrated in Figs. \ref{fig:5}(b)-\ref{fig:5}(d), where $E_{\mathrm{Bip}}$ is pushed towards zero for half-filled (and close to half-filled) bosonic media. Nevertheless, for $U_{BI}\gtrsim V_{B}$ the bound state emerges even in the CDW phase. Physically, this behavior can be understood by noting that the CDW exhibits a gap of order $\propto V_{B}$, boson-impurity interactions of comparable strength promote density fluctuations in the medium, thereby enhancing the induced impurity-impurity interaction. Due to the arise of the CDW, impurities bind more tightly for medium densities away from half-filling for $V_{B}>2t$. We have checked that when considering hard-core impurities, the binding energy is always positive (see Appendix \ref{Ap2}), highlighting the key role of impurity statistics on bipolaron formation. 

\begin{figure}[t!]
\centering
\includegraphics[width=1.0\columnwidth]{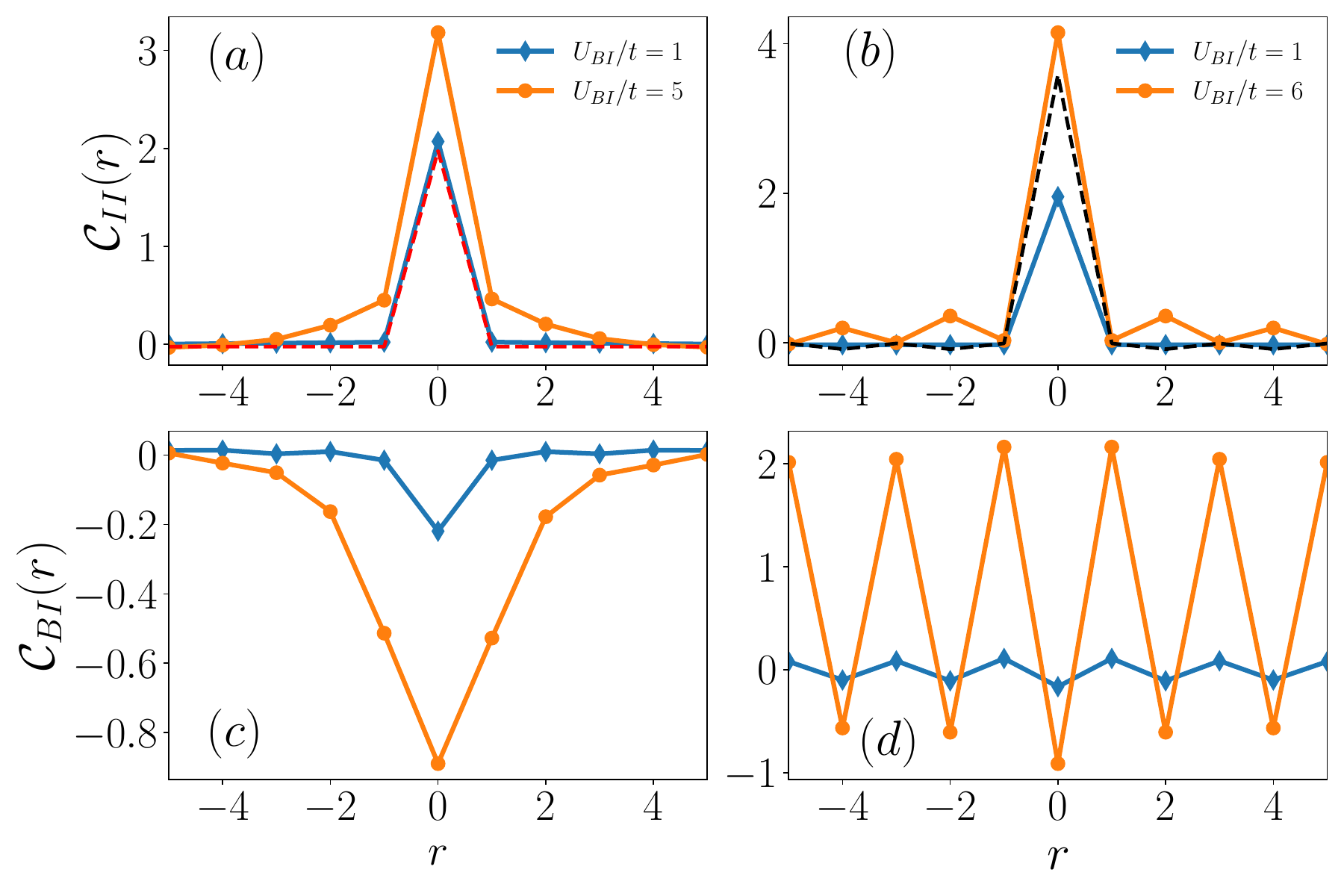}
\caption{(a) Impurity-impurity correlation $\mathcal{C}_{II}(r)$ as a function of the distance $r$ for $(n_{B},V_{B}/t)=(0.375,0)$, dashed line corresponds to the impurity-impurity correlation for two free impurities in the absence of the bosonic environment. (b) same as (a), but for $(n_{B},V_{B}/t)=(0.5,4)$, the dashed line is associated with the impurity-impurity correlation for two free impurities in the IHM. (c) Boson-impurity correlation $\mathcal{C}_{BI}(r)$ as a function of the relative distance $r$ for $(n_{B},V_{B}/t)=(0.375,0)$. (d) same as (c) but we consider $(n_{B},V_{B}/t)=(0.5,4)$. \label{fig:7}}
\end{figure}

Signatures of bound state formation can also be obtained by the impurity-impurity correlation function 
\begin{equation}
\mathcal{C}_{II}(i-j) = L[\langle\hat{n}_{I,j}\hat{n}_{I,i} \rangle -\langle\hat{n}_{I,j}\rangle\langle\hat{n}_{I,i}\rangle].
\end{equation}
As depicted in Fig. \ref{fig:7}(a), in the absence a of tight bound state $(V_{B}/t,U_{BI}/t)=(0,1)$, $\mathcal{C}_{II}(r)$ matches the impurity-impurity correlation function associated with two free impurities without bosonic environment (red dashed line). In contrast, for $(V_{B}/t,U_{BI}/t)=(0,5)$, the impurities correlate
$\mathcal{C}_{II}(r\neq0)\neq 0$ as a result of bipolaron formation. These correlations, absent in the free model, emerge from medium-induced interactions, highlighting the crucial role of the environment. 

Figure  \ref{fig:7}(b) illustrates $\mathcal{C}_{II}$ within the CDW phase $V_{B}/t=4$, weak boson-impurity interactions lead to similar correlations as that found within the superfluid phase. For $U_{BI}\gtrsim V_{B}$, the impurities promote density fluctuations on top of the CDW order, leading to bipolaron correlations that are a combination of both, binding correlations and out of phase CDW order. That is, due to the repulsive boson-impurity interaction, the bound state extends over the ordered hole positions of the bath. To show that the emerging correlations within the CDW phase are not entirely produced by the ionic Hubbard model modulations, we plot the impurity-impurity correlation function for two free impurities in the IHM (black dashed line). As can be seen, the free model is not able to reproduce the non-diagonal correlations.   

Alternatively to $\mathcal{C}_{II}$, the appearance of a bound state can be monitored through the boson-impurity correlation function $\mathcal{C}_{BI}$.  Figs. \ref{fig:7}(c) and \ref{fig:7}(d) display $\mathcal{C}_{BI}(r)$ within the SF and CDW phases of the bath, respectively. Blue curves correspond to unbound and orange to bound impurities. In the superfluid phase, the bound state manifests as a deeper and more widespread depletion of the medium density. This can be understood by noting that hole density of the medium is promoted as the impurities become correlated and stay close together. The spatial extent of this depletion matches the range of the impurity-impurity correlations. In the insulating phase, the hole density is ordered. Since it is energetically favorable for the impurities to bind at the hole positions, the bound state inherits the hole order, leading to a magnification of the alternating pattern in comparison with the unbound signal.

\section{Conclusions}
\label{CO}
Motivated by recent experimental advances with Rydberg-dressed gases in optical lattices~\cite{weckesser2024}, we consider a single and a pair of mobile impurities immersed in an interacting hard-core Bose gas in a one-dimensional lattice.  In particular, we investigate how the ground state properties of the impurities
change when the environment is in a superfluid or insulating charge density wave phase. In the superfluid phase, the impurity exhibits polaron-like behavior, forming a dressed quasiparticle. However, in the insulating phase, the impurity can retain its individual character, moving in a potential landscape shaped by the charge density wave order, which can be effectively described by an ionic Hubbard model. For the case of two impurities, we numerically demonstrate bound state formation without explicit impurity-impurity interactions and show its stability in the strongly interacting regime of the bath. Experimentally relevant density-density correlations also reveal clear signatures of this bound state. The discussed results provide valuable insights for ongoing lattice polaron experiments with ultracold gases in low-dimensional systems.

\section*{Acknowledgments}
We thank Arturo Camacho-Guardian and Thom\'as Fogarty for their valuable comments on our manuscript.
L.S. and G.A.D.-C. acknowledge support of the Deutsche Forschungsgemeinschaft (DFG, German Research Foundation)  under Germany’s Excellence Strategy, EXC-2123 QuantumFrontiers, Grant No.
390837967 and LAPA acknowledges financial support from PNRR MUR project PE0000023-NQSTI.

\appendix
\section{Single impurity}
\label{Ap1}

\begin{figure}[t!]
\centering
\includegraphics[width=1.0\columnwidth,height=0.78\columnwidth]{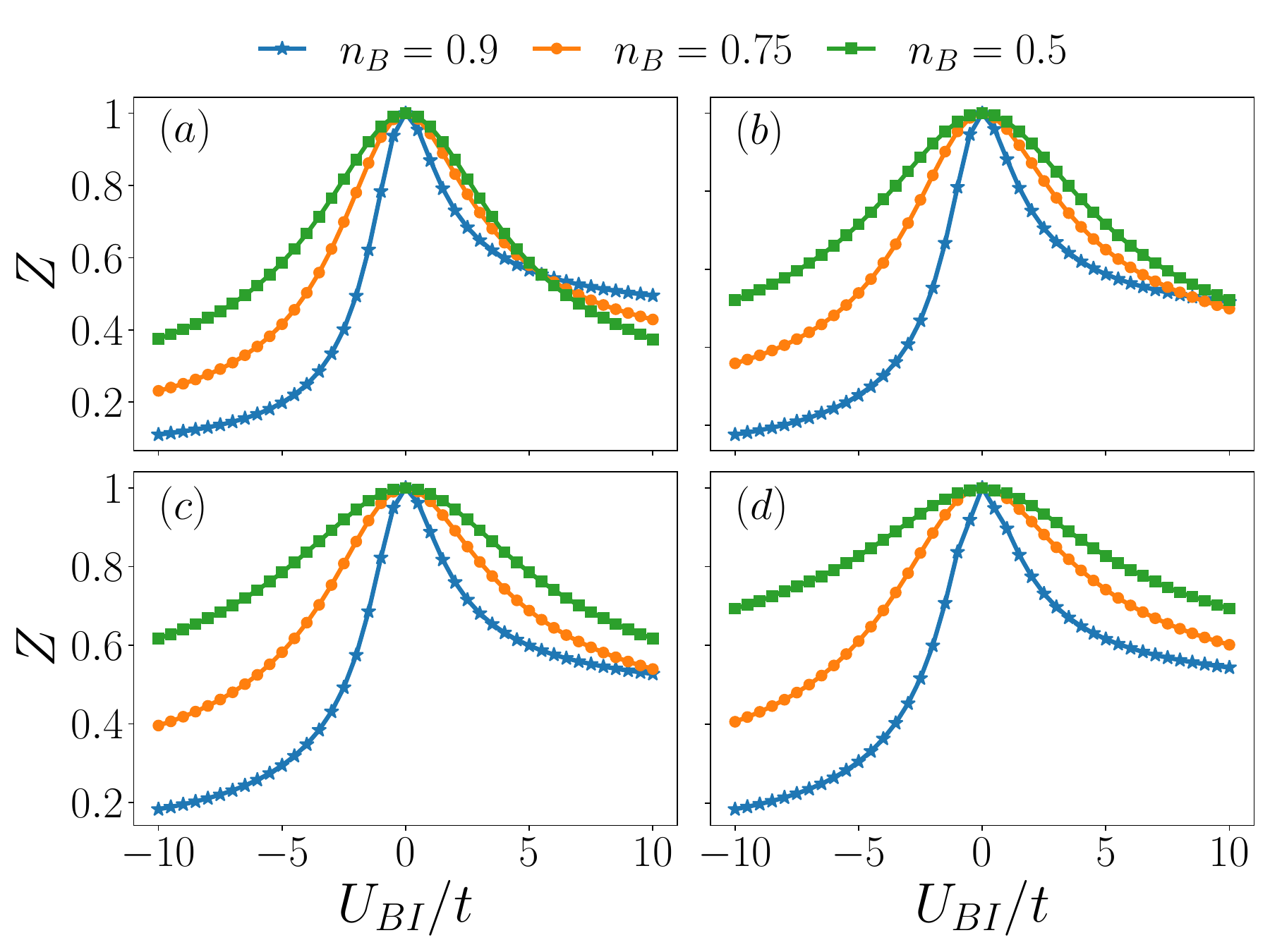}
\caption{Quasiparticle residue $Z$ as a function of the boson-impurity interaction $U_{BI}/t$ for three different values of the bosonic filling $n_{B}$. Panels (a)-(d) are associated with $V_{B}/t = 0,1,2$, and $6$, respectively.\label{fig:A2}}
\end{figure}

In this section, we provide extended results for the case of a single impurity. Due to the particle-hole symmetry $n_{B}\rightarrow 1-n_{B}$ of $H_{B}$, the Hamiltonian associated with the bath, the quasiparticle residue of the impurity inherits a symmetry under $(n_{B},U_{BI})\rightarrow (1-n_{B},-U_{BI})$. To show this behavior, in Fig. \ref{fig:A2} we 
plot $Z$ as a function of the boson-impurity interaction $U_{BI}/t$. In contrast to Fig. \ref{fig:3}, for $n_{B}>1/2$ the residue quickly decays when $U_{BI}/t$ becomes more negative. 

\begin{figure*}[t!]
\centering
\includegraphics[width=2.0\columnwidth]{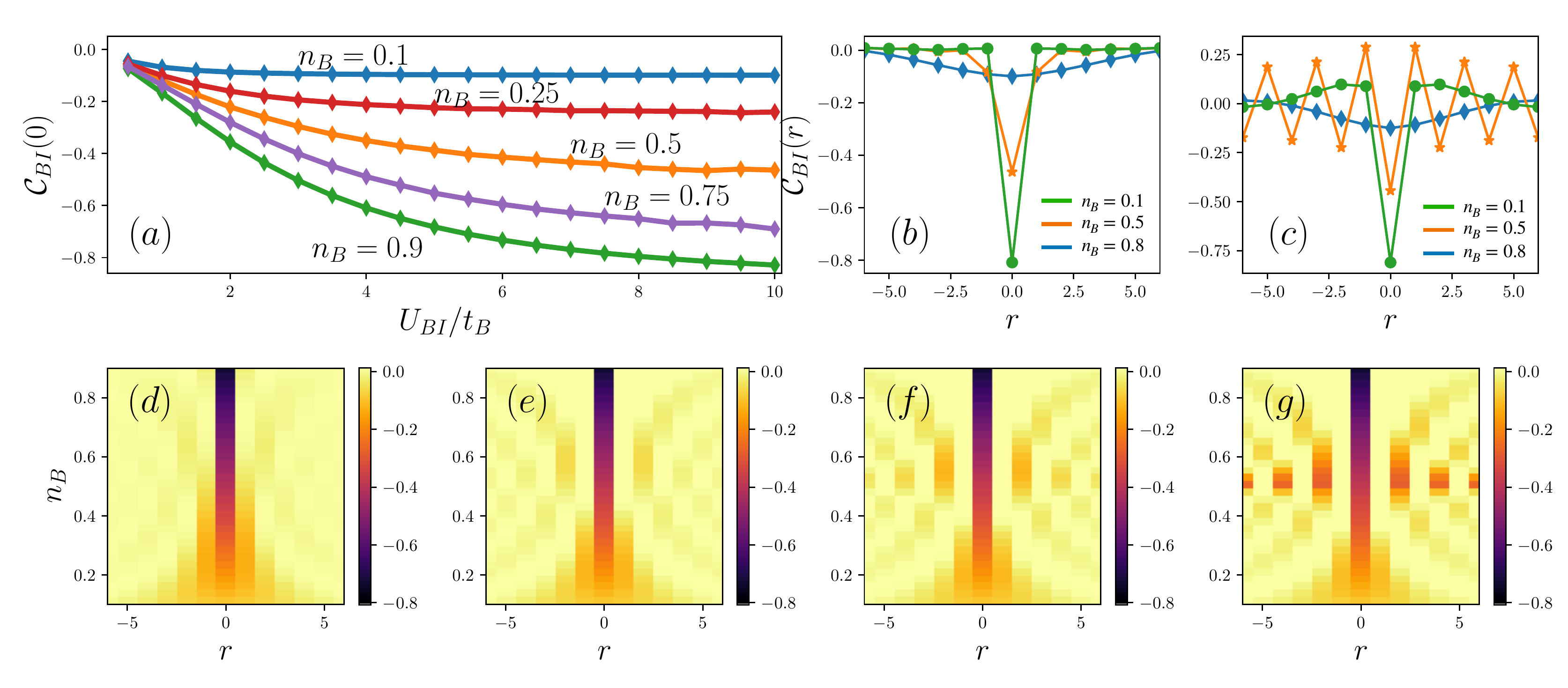}
\caption{(a) Density-density boson-impurity correlation $\mathcal{C}_{BI}(r)$ at the same site ($r=0$) as a function of the interaction $U_{BI}/t_{B}$ for several bosonic filling factors $n_{B}$ and fixed $V_{B}=0$. (b) $\mathcal{C}_{BI}(r)$ as a function of the relative distance $r$ for three different values of $n_{B}$ and $(V_{B}/t,U_{BI}/t)=(0,10)$. (c) Sames as panel (b) but for $(V_{B}/t,U_{BI}/t)=(4,10)$. (d)-(g) Development of the density-density boson-impurity correlations as the medium transitions from the superfluid to the charge density wave for $U_{BI}/t=6$. Panels (d), (e), (f), and (g) are associated with $V_{B}/t=0,1,2$ and $4$, respectively.   \label{fig:A3}}
\end{figure*}

To end this section, we present extended results of the boson-impurity correlations for $U_{BI}/t>0$. In contrast to Fig. \ref{fig:4}, for $U_{BI}/t>0$ density fluctuations of the bosonic medium increase as the $n_{B}$ increases. Figs. \ref{fig:A3}(d)-\ref{fig:A3}(g) display in a color scheme the change of $\mathcal{C}_{BI}(r)$ as the medium transitions from the superfluid to the charge density wave phase, we consider $U_{BI}/t=6$. As for the attractive case, the impurity provides an unambiguous probe of the emerging correlations of the medium as it enters the CDW phase.

\section{Two impurities}
\label{Ap2}

\begin{figure}[t!]
\centering
\includegraphics[width=1.0\columnwidth]{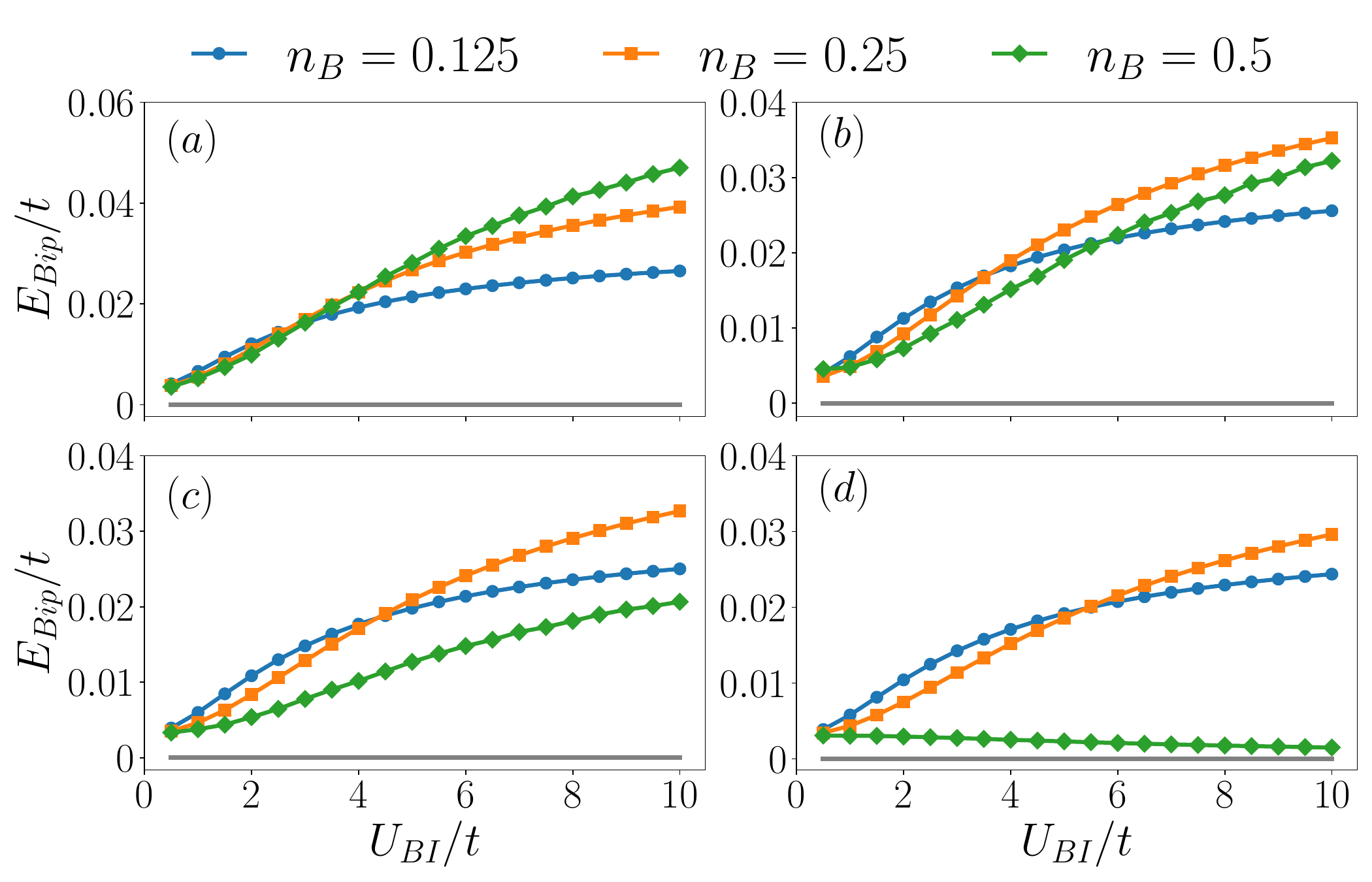}
\caption{Binding energy $E_{\mathrm{Bip}}$ of two hard-core impurities as a function of the boson-impurity interaction $U_{BI}/t$ for several values of the bosonic filling $n_{B}$. Panels (a), (b), (c), and (d) consider $V_{B}/t=0,1,2$, and $3$, respectively. The gray line represents a guide for the eye for zero energy.\label{fig:A4}}
\end{figure}

In the following, we provide extended results associated with the case of two impurities. In contrast to the case discussed in the main manuscript, we now consider hard-core impurities, i.e., we impose the double occupancy restriction on the impurities $(a_{i}^{\dagger})^{2}=0$. Under this condition, the impurities behave as spinless fermions, and consequently, direct contact interactions between the impurities are forbidden. Figure \ref{fig:A4} shows the binding energy of two hard-core impurities
as a function of the boson-impurity coupling for fixed bath density. In stark contrast to soft-core impurities (see Fig. \ref{fig:5}), $E_{\mathrm{Bip}}$ remains non-negative across the investigated range of $U_{BI}/t$, indicating the absence of bound state formation. Our results suggest that in the considered model, induced contact interactions play the main role in bipolaron formation, as hard-core impurities do not exhibit two-body bound states. Nevertheless, it is important to note that previous studies have shown that induced nonlocal interactions in soft-core Bose baths can lead to binding of two spinless Fermi impurities confined in harmonic traps \cite{PhysRevB.100.245419}. 
\newpage

\bibliography{Bibliography}

\begin{thebibliography}{79}
\bibitem [{not()}]{note1}%
  \BibitemOpen
  \href@noop {} {}\bibinfo {note} {In the thermodynamic limit, Eq.~(6) reads
  $E_{\text{Bip}} = E_2 - 2E_1 + E_0 = 0$ when the polarons are far from each
  other $|R_1 - R_2| \to \infty$. The finite-size correction depends on
  $\delta$, which defines our numerical zero. In this case, we find
  $E_{\text{Bip}} = E_2 - 2E_1 + E_0 = \delta t$ as $|R_1 - R_2| \to
  \infty$}\BibitemShut {NoStop}%
\end{thebibliography}
\end{document}